\def\etal{{et al. }}
\def\lum{{\rm erg \, s^{-1}} }

\def\chisq{$\chi^2$}

\documentstyle[12pt,aaspp4,epsf]{article}
\begin{document}
\lefthead{Donahue}
\righthead{A Very Hot, High Redshift Cluster}
\slugcomment{Submitted to \apj, June, 1997}
\title{A Very Hot, High Redshift Cluster of Galaxies: \\
More Trouble for $\Omega_0=1$}
\author{Megan Donahue and G. Mark Voit} 
\affil{Space Telescope Science Institute \\3700 San Martin Drive \\
Baltimore, MD 21218 \\ donahue@stsci.edu, voit@stsci.edu}
\author{Isabella Gioia\altaffilmark{1,}\altaffilmark{2} and 
Gerry Luppino\altaffilmark{2}}
\altaffiltext{1}{Home Institution: 
Istituto di Radioastronomia del CNR, Via Gobetti,
101 40129 Bologna, Italy}
\altaffiltext{2}{Visiting Astronomer at CFHT,
operated by the National Research Council of Canada, le Centre
National de la Recherche Scientifique de France and the University
of Hawaii, and at the W. M. Keck Observatory, jointly
operated by the  California Institute of Technology and the
University of California.}
\affil{Institute for Astronomy \\ 2680 Woodlawn Drive \\ Honolulu, HI 96822 
\\ gioia@galileo.ifa.hawaii.edu, ger@galileo.ifa.hawaii.edu}
\author{John P. Hughes}
\affil{Department of Physics and Astronomy\\ Rutgers University \\ PO Box 849 \\
Piscataway, NJ 08855-0849 \\ jackph@physics.rutgers.edu}
\author{John T. Stocke}
\affil{University of Colorado \\ Center for Astrophysics and Space
Astronomy \\ CB 389 \\ Boulder CO 80309 \\ 
stocke@casa.colorado.edu}

\newpage

\begin{abstract}
We have observed the most distant ($=0.829$) cluster of galaxies in the 
{\em Einstein} Extended Medium Sensitivity Survey (EMSS), with the 
ASCA and ROSAT satellites.  We find an X-ray temperature of 
$12.3^{+3.1}_{-2.2} \, {\rm keV}$ for this cluster, and the ROSAT map
reveals significant substructure.  The high temperature of MS1054-0321
is consistent with both its approximate velocity dispersion, based on 
the redshifts of 12 cluster members we have obtained at the Keck and 
the Canada-France-Hawaii telescopes, and with its weak lensing 
signature.  The X-ray temperature of this cluster implies
a virial mass $\sim 7.4 \times 10^{14} \, h^{-1} \, M_\odot$,
if the mean matter density in the universe equals the critical 
value ($\Omega_0 = 1$), or larger if $\Omega_0 < 1$.  
Finding such a hot, massive cluster in the EMSS is extremely
improbable if clusters grew from Gaussian perturbations in an
$\Omega_0 = 1$ universe.  Combining the assumptions that $\Omega_0 = 1$
and that the intial perturbations were Gaussian with the
observed X-ray temperature function at low redshift, we show
that this probability of this cluster occurring in the volume 
sampled by the EMSS 
is less than a few times $10^{-5}$.
Nor is MS1054-0321 the only hot cluster at high redshift;
the only two other $z > 0.5$ EMSS clusters already observed with
ASCA also have temperatures exceeding 8~keV.  Assuming again that
the initial perturbations were Gaussian and $\Omega_0 =1$, we
find that each one is improbable at the $< 10^{-2}$ level. 
These observations,
along with the fact that these luminosities and temperatures 
of the high-$z$ clusters all agree 
with the low-$z$ $L_X-T_X$ relation, argue strongly that
$\Omega_0 < 1$.  Otherwise, the initial perturbations must
be non-Gaussian, if these clusters' temperatures do indeed
reflect their gravitational potentials.

\end{abstract}

\section{Introduction}

Clusters of galaxies occupy a special position among 
celestial self-gravitating structures, as they are the 
largest virialized objects in the universe. If all the 
cosmic structures we currently see grew hierarchically from 
small seed perturbations in the early universe, as many
lines of evidence now suggest (e.g. Peebles 1993),
then the largest clusters of galaxies are also the most 
recent virialized structures to have appeared on the
cosmological scene.  The most massive of all clusters, which
represent the most extreme high-density peaks on scales
$\sim 10^{15} \, M_{\odot}$ in the initial field of density
perturbations, correspond to several-$\sigma$ events
if the perturbations are Gaussian. 

Because massive clusters represent the high-$\sigma$ 
tail of the distribution, their number density 
should rise dramatically as the overall amplitude of 
the perturbations grows.  This characteristic of cluster formation 
renders massive cluster evolution particularly sensitive to 
the mean matter density of the universe, currently equal 
to $\Omega_0$, in units of the critical density $3 H_0^2 / 8 \pi G$.  
If $\Omega_0 \approx 1$, perturbations should still grow like
$(1+z)^{-1}$, and evolution in the number density of distant
clusters with redshift 
should be rapid.  If $\Omega_0 \ll 1$, perturbation growth
has virtually stopped, and massive cluster evolution should be
much more mild.

Evolution in the distribution of cluster masses sensitively
indicates $\Omega_0$,
but measuring the virial masses of clusters ($M_{\rm vir}$) is 
difficult, even at low redshifts.  A much more straightforward 
observational task is to measure how the X-ray temperatures 
($T_X$) of clusters depend on redshift and then to use a model 
to relate $T_X$ to $M_{\rm vir}$ (e.g. Oukbir \& Blanchard 1992). 
The {\em Advanced Satellite for Cosmology and Astrophysics} (ASCA) 
(Tanaka, Inoue, \& Holt 1994) now makes such tests possible because 
it can measure the temperatures of the most luminous
X-ray clusters at high redshifts (Yamashita 1994; Furuzawa \etal
1994; Donahue 1996). Henry (1997) has recently published a temperature
function for 10 massive clusters at $z \sim 0.3$, chosen from the 
{\em Einstein} Extended  Medium Sensitivity Survey (EMSS) (Gioia \etal 1990;
Henry \etal 1992).  His comparision of the $z \approx 0.3-0.4$ 
temperature function with the $z \approx 0$ temperature function 
suggests that $\Omega_0 \approx 0.5$ and rules out $\Omega_0 = 1$ 
with 99\% confidence.

In this paper we attempt to extract the maximum leverage out of
this test for $\Omega_0$ by measuring the temperature of the 
highest-redshift cluster in the EMSS, MS1054-0321 at $z = 0.83$.
We present both an ASCA spectrum and a ROSAT HRI image
of this very luminous cluster ($L_x = 5.6 \times 10^{44} \, h^{-2} 
\, \lum$ 2-10 keV rest frame, $L_x = 1.1 \times 10^{45} \, h^{-2} \, 
\lum$ bolometric). For $q_0=0.1$, $L_x = 7.7  \times 10^{44} \, h^{-2} 
\, \lum$ 2-10 keV rest frame.   
Luppino \& Kaiser (1997) provide a true color optical
image and a weak lensing map.  
After presenting the observations in \S~2, we examine how they 
constrain $\Omega_0$ in \S~3, finding that they strongly indicate
$\Omega_0 < 1$.  Section~4 summarizes our results.
The ASCA and ROSAT observations 
we present here are part of a larger program to measure the 
X-ray temperatures of three high-$z$, X-ray luminous clusters 
in the EMSS, the other two being MS0451-03 at $z=0.53$ (Donahue 1996) 
and MS1137+66 at $z=0.78$ (Donahue 1998, in preparation). Throughout
the paper, we parametrize $H_0 = 100h$ km/sec/Mpc and assume $q_0 = 0.5$,
except where noted.

\section{Observations and Data Reduction}

This section describes the data indicating that MS1054-0321 is
an unusually massive cluster.  We first discuss the ASCA spectra
showing that MS1054-0321 has a temperature in excess of 10~keV.
Next we introduce the optical velocity dispersion and lensing
data on this cluster, which agree with its high X-ray temperature.
Then we show a ROSAT HRI image that reveals substructure
in this cluster, suggesting that it may not be fully relaxed.

\subsection{ASCA Spectra}

ASCA executed a single long ($\sim 70,000$ s) exposure
of the cluster MS1054-0321 during May 23-25, 1995.  
This satellite carries four independent X-ray telescopes,
two with Gas Imaging Spectrometers (GIS) and two with
Solid-State Imaging Spectrometers (SIS), from which we
obtained four independent datasets.
The GIS detectors are imaging gas scintillation proportional 
counters with energy resolutions of 8\% at 5.9 KeV and 
50\arcmin~ diameter fields of view.
The two SIS detectors are X-ray sensitive CCDs 
with energy resolutions of 2\% at 5.9 KeV and
22\arcmin~ by 22\arcmin~ square fields of view.  
These observations employed the SIS detectors in 2-CCD mode.
However, we restricted the data analysis to the CCD containing the
target and only used the second CCD to check the background estimates.

To prepare the data for analysis, we extracted clean 
X-ray event lists using a magnetic cut-off rigidity 
threshold of 6 GeV~c$^{-1}$ and the recommended minimum elevation 
angles and bright Earth angles to reject background contamination 
(see Day et al. 1995). 
Events were extracted from within circular apertures of radii  
3.25 arcmin (SIS1), 2.53 arcmin (SIS0), and 5.0 arcmin (GIS) 
to maximize the ratio of signal to background noise.  
We rejected $\sim$98\% of cosmic ray events by using only SIS 
chip data grades of 0, 2, 3, 4, and we rejected hot and 
flickering pixels. Light curves for each instrument were 
visually inspected, and time intervals with high background 
or data dropouts were excluded manually.
We rebinned the SIS data in the standard way to 512 spectral 
channels using Bright2Linear (see Day \etal 1995) with the lowest
13 channels flagged as bad, and we regrouped the GIS data so 
that no energy bin had fewer than 25 counts. 
Table~\ref{asca} gives the resulting net count rates
and effective exposure times. We do not expect the derived
X-ray fluxes to be consistent between the detectors because
the target was not centered and some of the extended flux may 
have missed the SIS detectors.

Background estimates were taken from the regions of the detector
surrounding the cluster detection, from summed deep background 
images supplied by the ASCA Guest Observer Facility (GOF), and from 
the second CCD.  The deep background spectra were extracted with 
the same spatial and background removal filters used to extract 
the source data.  The intensities of the local background and the 
deep background were consistent with each other, but the uncertainties 
for the deep background estimate were larger (because the source 
extraction area was smaller). We therefore used the local background 
estimates for our analysis.

We extracted SIS spectra binned in energy channels in two different
ways: (1) binning by a factor 4 in energy, and (2) regrouping the data
into energy bins containing a minimum of 16 counts. The first procedure 
provided slightly better constraints on iron abundance because it 
maintained some of the spectral resolution around the 6.7 keV
Fe line, preserving the predicted contrast between the counts 
in the energy bin where the line should be and the continuum channels.
The second procedure yielded slightly better contraints on temperature.
After rebinning the data, we restricted our fitting of the SIS spectra
to the $0.5-5.5$~keV range over which the signal-to-noise was adequate.
The adequate range of the GIS spectra extended to 7~keV. 

To analyze the spectra, we used XSPEC (v10.0) from the software 
package XANADU available from the ASCA GOF (Arnaud 1996). 
We fitted the spectral data from the four ASCA datasets and their 
respective response files (see Day \etal 1995).  The individual
SIS response matrices were generated with the tool {\em sisrmg} 
(1997 version), which takes into account temporal variations 
in the gain and removes the inconsistencies seen in data analyzed 
with the standard SIS response matrices.

\begin{figure}
\plotone{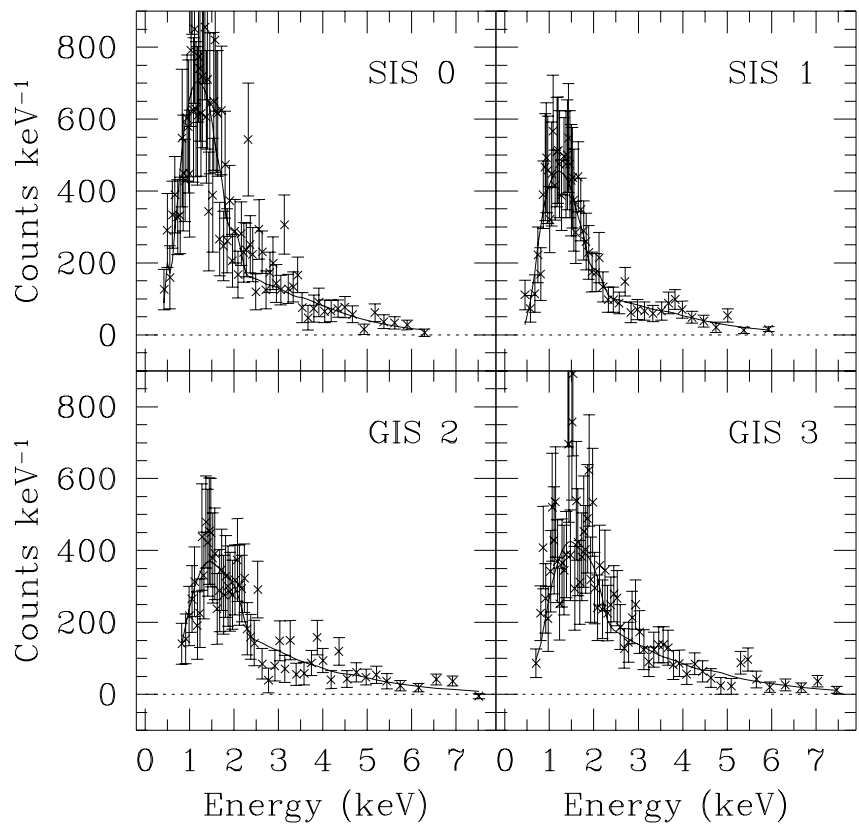}
\figcaption[fig1.ps]{We plot the X-ray spectra from each of
the four ASCA telescopes for the cluster MS1054-0321. 
Each spectrum has over 1000 X-ray
counts. (See Table 1.) The best-fit model is drawn with a 
solid line. The spectra we display were binned such that
each energy bin contained a minimum of 16 (SIS) or 25 (GIS) counts. \label{spectra}}
\end{figure}

\begin{figure}
\plotfiddle{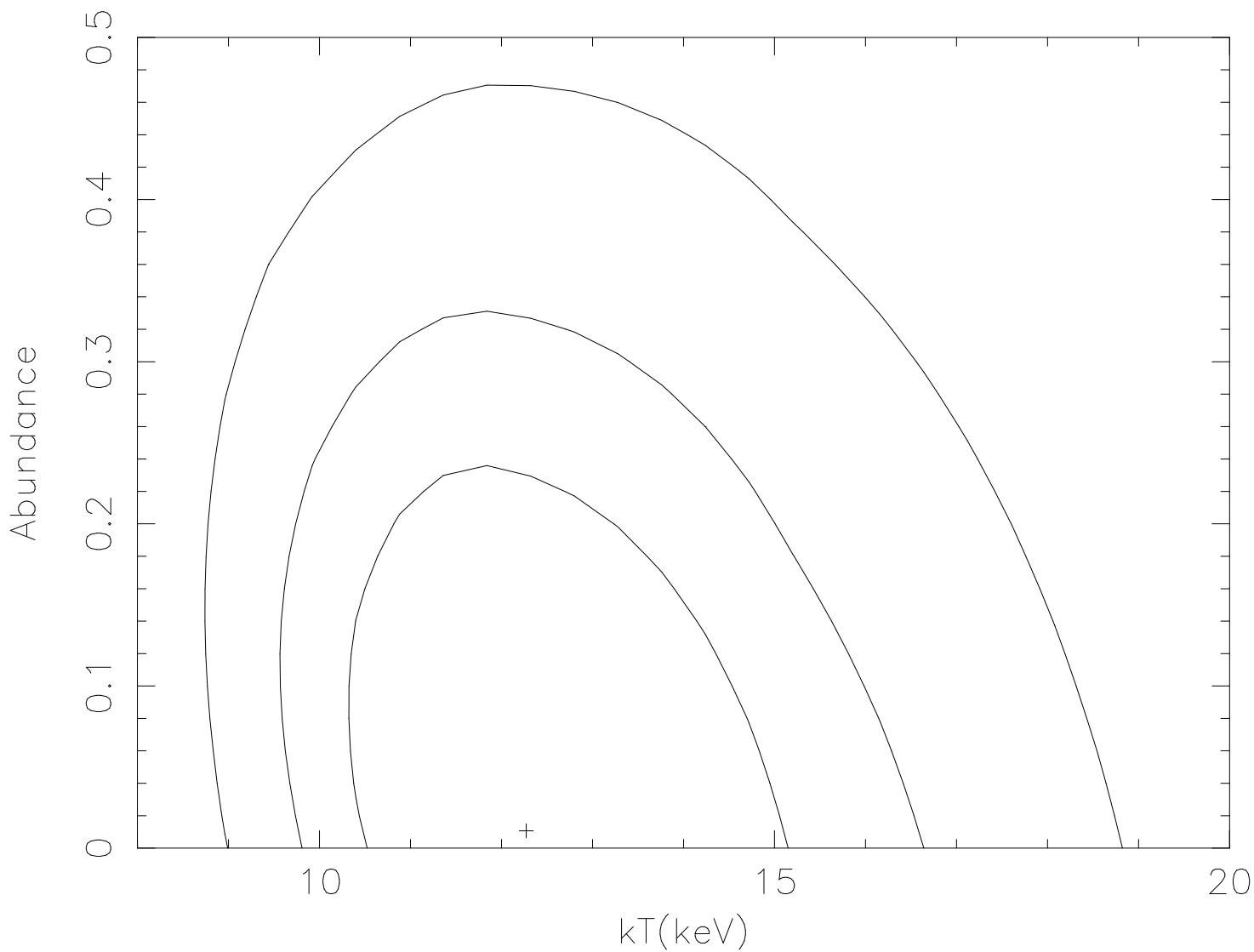}{6in}{-90}{80}{80}{-320}{500}
\figcaption[fig2.ps]{We plot the two-dimensional 
$\chi^2$ contours at 68.3\%, 
90\% and 99\% confidence levels ($\Delta \chi^2 = 2.30$, 4.61 and 9.21) 
for the cluster iron abundance in units of the solar abundance 
and temperature in keV. These contours represent the confidence
contours of simultaneous fits to 
the SIS spectra  and to GIS spectra
binned such that each bin contained a minimum of 
16 and 25 counts respectively. \label{contour}}
\end{figure}

The standard model we used to fit the data was a Raymond-Smith 
thermal plasma with a temperature $T_X$, absorbed by cold Galactic 
gas with a characteristic column density of neutral hydrogen $N_{\rm H}$.
We did not detect a significant intrinsic absorption component, 
to a 90\% upper limit of $N_H < 1.4 \times 10^{21}$ cm$^{-2}$. 
Thus, we fixed soft X-ray absorption at an HI value of
$3.7 \times 10^{20}$ cm$^{-2}$ (Gioia \etal 1990).  
Each SIS spectrum was fit with its own normalization; 
the normalizations of the GIS spectra were constrained to be the same.  
The parameters varied to provide a fit to 4 spectra were therefore
the 3 independent normalizations (1 for each SIS spectrum, 1 for 
both GIS spectra), a single emission-weighted temperature, 
and the metallicity (iron abundance). Examples of the binned 
spectra and best-fit model convolved with each telescope's 
response matrix are found in Figure~\ref{spectra}. 
The best-fit redshifted
temperature to the binned-by-four spectrum was $12.4 \pm ^{3.6}_{2.3}$keV 
with an upper limit on iron abundance of $0.22$ solar, keeping the 
absorption fixed. If we used the SIS spectra regrouped to a minimum
of 16 counts per bin, the best-fit temperature was virtually the
same: $12.3 \pm ^{3.1}_{2.1}$ keV with an upper limit on the iron 
abundance of $<0.25$ solar.  All uncertainties quoted are the 90\% 
confidence levels for a 1-dimensional fit ($\Delta \chi^2 = 2.70$), 
and all fits have an acceptable reduced  \chisq~ of $\sim0.8-0.9$. 
Figure~\ref{contour} shows the two-parameter $\chi^2$ contours 
for the cluster metallicity and $T_X$ in keV.

\subsection{Optical Corroboration}

The available optical data on this cluster corroborate the
high temperature measured by ASCA.  Weak lensing observations of
MS1054-0321 by Luppino \& Kaiser (1997) show that its mass within
$0.5 \, h^{-1} \, {\rm Mpc}$ is $(5-11) \times 10^{14} \, h^{-1} \,
{\rm Mpc}$, depending on the redshift distribution of the lensed
galaxies.  In an isothermal potential, its mass distribution 
corresponds to an isotropic one-dimensional velocity dispersion 
$\sim 1100-2200 \, {\rm km \, s^{-1}}$, consistent with the value 
expected from the ASCA temperature: $(kT_X/\mu m_p)^{1/2} = 
1400 \pm 170$ km s$^{-1}$ (90\% confidence limits).

In addition, we have obtained the redshifts of twelve cluster members 
with the CFHT Multi-Object Spectrograph (MOS) on April 13, 1994 and 
with the LRIS (Oke et al. 1995) at Keck on January 4, 1995. These are listed in 
Table~\ref{Redshifts}. Using Timothy Beers' program ROSTAT, we 
calculated their mean redshift and velocity dispersion, finding  
$z = 0.829 \pm 0.008$ and a velocity dispersion of $1360 \pm 450 
\, {\rm km \, s^{-1}}$ (90\% confidence interval).  (The observed 
redshift dispersion has been corrected to first order for cosmological 
redshift by dividing by $1+z$). Several alternative methods (Beers, 
Flynn, \& Gebhardt 1990) give the same answer within the uncertainties.  
While the small numbers of redshifts do not constrain the velocity 
dispersion very well, the value we find is consistent with both the 
lensing results and the X-ray temperature.

\begin{figure}
\plotone{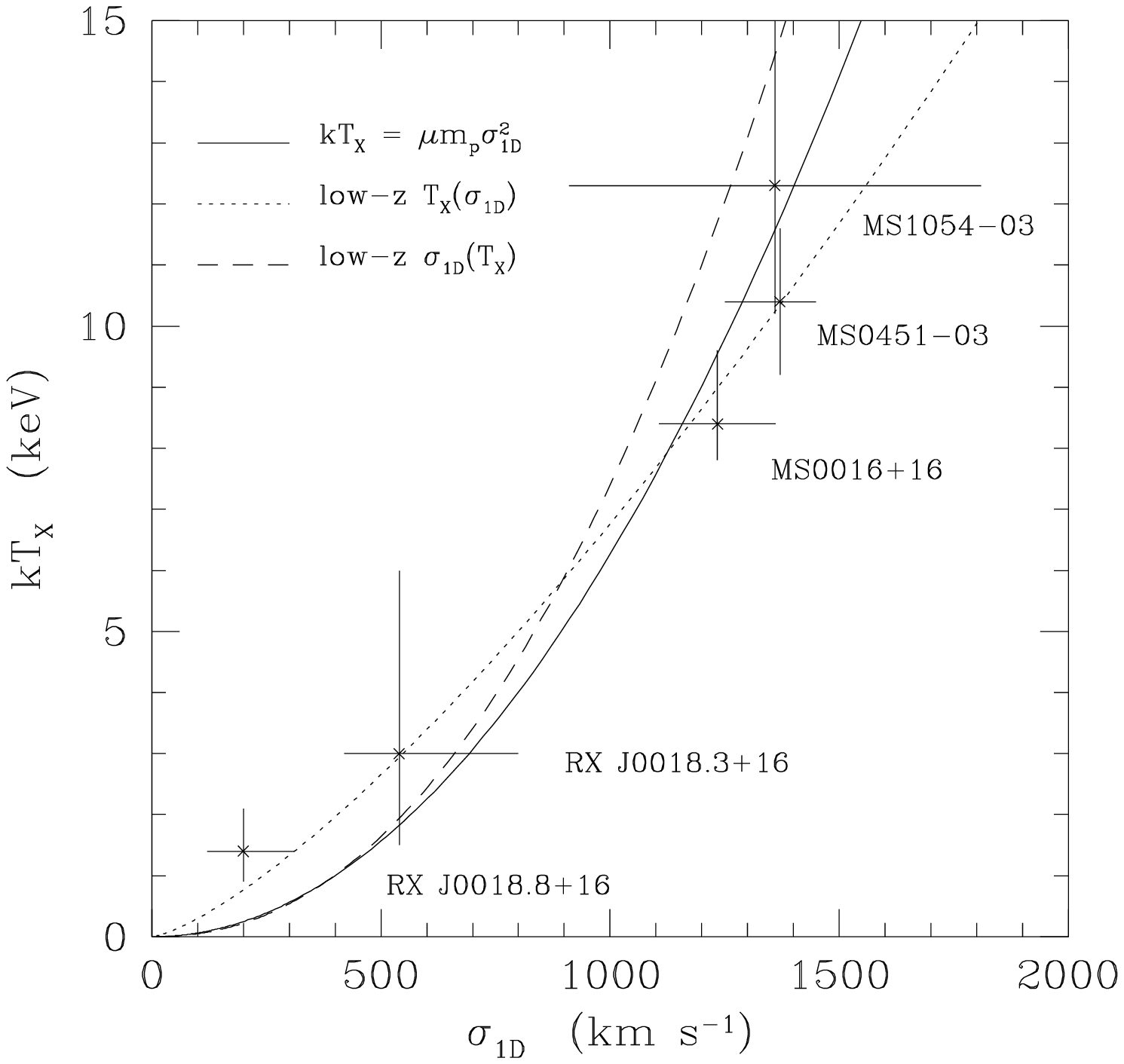}
\figcaption[fig3.ps]
{The relation between line-of-sight velocity dispersion ($\sigma_{\rm 1D}$)
and X-ray temperature ($T_X$) for several clusters at $z>0.5$.  
The solid lines illustrates the relation $kT_X = \mu m_p \sigma_{\rm 1D}^2$,
and the other two lines show the $\sigma_{\rm 1D}$-$T_X$ correlations 
derived from low-$z$ clusters by Edge \& Stewart (1991).  The dotted
line gives the correlation when $T_X$ is the dependent variable, and
the dashed line gives the correlation when $\sigma_{\rm 1D}$ is the
dependent variable.  Temperatures of high-$z$ clusters at $kT_X \sim 10
\, {\rm keV}$ appear to indicate the depths of their gravitational 
potentials with respectable accuracy. \label{tx_sigma}}
\end{figure}

Recent observations of other hot, high-$z$ clusters further
support the notion that their temperatures faithfully reflect
their gravitational potentials.  Figure~\ref{tx_sigma} shows the relationship
between the X-ray temperatures and velocity dispersions of
three hot X-ray clusters at $z > 0.5$ from the EMSS and
two cooler clusters that are companions of MS0016+06 at $z = 0.54$ 
(Hughes, Birkinshaw, \& Huchra 1995; Hughes \& Birkinshaw 1998).
The X-ray temperatures for the three hot clusters are from this paper, 
Yamashita (1994), Furuzawa \etal (1994), and Donahue (1996), and their 
velocity dispersions are from this paper and Carlberg \etal (1996).
The data for the two cooler clusters are from Hughes \etal (1995)
and Hughes \& Birkinshaw (1998).  The hotter clusters agree
with the relation $kT_X = \mu m_p \sigma_{\rm 1D}^2$, while
the cooler clusters might be slightly hot for their velocity
dispersions, but nevertheless in agreement with the low-$z$
$\sigma$-$T_X$ relation of Edge \& Stewart (1991a).
 
\subsection{X-Ray Imaging}

Our ROSAT data on this cluster suggest that we ought to interpret
its temperature and velocity dispersion cautiously.
We observed MS1054-0321 with the ROSAT HRI for over 121,590 
seconds on December 6-18, 1996, yielding $1060 \pm 90$ net counts
(including an earlier 12,136 second observation on 
May 5-6, 1996.)  Figure~\ref{plate} (Plate 1) shows a map of X-ray contours 
overlaid on a ground-based optical image of the cluster. For this 
figure, the X-ray emission was smoothed adaptively with Gaussians chosen
such that the product of the number of counts and $\sigma^2$ was roughly
constant. The minimum  
contour corresponds to $1$ count per $8\arcsec 
\times 8\arcsec$, which is $\sim 3-\sigma$. 
The cluster emission is clearly extended over a scale of at least one
arcminute. With the spatial resolution of the HRI, we have
resolved the X-ray emission into two or three clumps and an extended component, 
clearly indicating that this cluster is not regular. 

King models centered on 
the cD galaxy in MS1054-0321 tend to have high $\beta$ values 
($\beta \sim 0.7-1.0$) and large core radii ($\sim$ 1' or 
$\sim 500 h_{50}^{-1}$ kpc), although these values are not 
well constrained by the HRI data because of the cluster's 
irregular X-ray shape and clumpiness. A King model is rejected
by the data, with reduced $\chi^2 > 2$ for both circular and
elliptical King models.
  
The most
western peak is the most significant, containing $\sim 144 \pm 25$
counts within 24\arcsec~ above the smooth cluster component. 
The central peak is smaller, with $\sim 90
\pm 24$ counts and the easternmost peak is the least significant
at $44\pm 14$ counts above the smooth component. We made a basic
statistical test for bimodality of the X-ray distribution of photons
with the Lee statistic (Fitchett 1988; Lee 1979).
We calibrated the Lee statistic by producing multiple simulations of the 
null hypothesis in which the cluster emission is distributed in a 
King model ($\propto (r^2 + r_{core}^2)^{-3\beta + 1/2}$) where the
core radius is 56\arcsec~ and $\beta=2/3$ or $1$. We found that the 
cluster is at least bimodal at greater than a 99.9\% confidence level for
either null hypothesis.

The peaks in the X-ray emission do not coincide with any particular 
optical feature to within the 6~arcsec aspect uncertainty of the
X-ray image. The cluster is elongated 
east-west, in the direction of the galaxy distribution, and a tail 
extends south of the cluster.  Its X-ray emission roughly coincides 
with the dark matter contours as plotted by Luppino \& Kaiser (1997), 
including the east-west and southerly extensions.  The spatial 
resolution of the weak lensing map is not sufficient to confirm 
the sub-clumps seen in the X-ray map. The cluster X-ray emission 
also coincides with the shape of the Sunyaev-Zeldovich temperature 
decrement detected in centimeter interferometry experiments by 
L. Grego (PhD thesis, in preparation).  

The asymmetric appearance of the cluster in these high resolution 
images warns us that the relationship between its mass and its 
temperature might not be as simple as a spherical, hydrostatic,
isothermal model would predict.  However, recent hydrodynamic 
simulations of cluster formation beginning with cosmological 
initial conditions indicate that the systematic errors on cluster 
masses derived using a scaling relation between mass and temperature
are relatively small, typically $< 20$\% even when some substructure
is present (Evrard, Metzler, \& Navarro 1996).  Higher resolution 
simulations of individual cluster mergers, whose initial conditions
are not dictated by hierarchical structure formation, indicate 
that errors in the estimated mass can be somewhat larger 
($\lesssim 50$\%) after a recent merger, but these errors 
tend to yield {\em underestimates} of the cluster's mass 
(Roettiger, Burns \& Loken 1996; Schindler 1996).
The following discussion will therefore focus on the implications 
of the cluster temperature alone, assuming that the emission-weighted 
temperature is representative of the virial temperature.  
Mass errors at the 20\% level do not affect the qualitative results 
of the ensuing discussion. 

\section{Constraints on Temperature Evolution}

This cluster, MS1054-0321, is one of the hottest clusters known. 
If $\Omega_0 = 1$, structure formation simulations show that we 
can estimate its virial mass by assuming that the mean density
within its virialized region is $\sim 200$ times the critical 
density at the cluster's redshift and that the cluster itself 
is isothermal (Evrard \etal 1996; Eke \etal 1996). 
The virial mass of MS1054-0321 would then be approximately 
$M_{\rm vir} \sim [2kT_X/\mu m_p (1+z)]^{3/2} (10H_0G)^{-1}$ 
$\sim 7.4 \times 10^{14} \, h^{-1} \, M_{\odot}$ within $r_{200} =
1.5 h^{-1}$ Mpc. (We note that this relation is equivalent 
to assuming $\beta = 1$ in a standard X-ray mass relation for
an isothermal gas.) 
Such a massive cluster at such a high redshift severely challenges
models of hierarchical structure formation with $\Omega_0 = 1$ 
(Evrard 1989; Peebles, Daly, \& Juszkiewicz 1989; Donahue 1993).
This section elucidates how severe those challenges are.

Because the virial masses of clusters are much harder to measure
than their temperatures, we will restrict this analysis to
focus on temperature evolution, relying on $T_X$ to be a surrogate
for $M_{\rm vir}$.  The analysis will therefore depend on the
temperature-mass relation we assume.  In an $\Omega_0 = 1$
universe, this relation is relatively simple:  $T_X \propto 
M_{\rm vir}^{2/3} (1+z)$.  If $\Omega_0 \ll 1$, this relationship 
breaks down below $z \sim \Omega_0^{-1} - 1$ as the development
of structure stagnates.  

\subsection{Rarity of Hot Clusters at High-$z$}

In order to determine the rarity of a cluster as hot as MS1054-0321
in an $\Omega_0 = 1$ universe seeded with Gaussian perturbations,
we will assume that the Press-Schechter formula (Press \& Schechter
1974) adequately describes the mass function of virialized objects,
an assumption borne out by numerical simulations (e.g. Lacey \& Cole
1994, Eke \etal 1996).
In integral form, the comoving mass density of virialized objects
with virial masses greater than $M$ is then
\begin{equation}
\rho(>M) = \rho_0 \, {\rm erfc}(\nu_c/\sqrt{2}) = \frac{2\rho_0}{\sqrt{\pi}} \int^{\infty}_{\nu/\sqrt{2}} e^{-x^2} \, dx \; \; ,
\end{equation}
where $\rho_0$ is the current mean matter density and $\nu_c$ is the
critical threshold at which perturbations virialize.  This threshold
is $\nu_c = \delta_c(t) / \sigma(M) D(t)$, where $\sigma(M)$
describes the rms amplitude of linear perturbations on mass scale $M$,
$D(t)$ describes their linear growth rate, and $\delta_c(t)$ is the
linear overdensity at which perturbations virialize.  If $\Omega_0 = 1$,
then $\nu_c \propto (1+z)$ at a fixed mass scale $M$.

This integral form of the Press-Schechter formula makes its theoretical
underpinnings somewhat more apparent than the more familiar differential
form does.  The formalism assumes that the original linear density 
perturbations from which structure grew, smoothed over a length scale
corresponding to mass $M$, follow a gaussian distribution with a 
dispersion proportional to $\sigma(M)$.  As the perturbations grow, 
the most prominent of them collapse and virialize when their amplitudes 
exceed some threshold.  The parameter $\nu_c$ expresses how many standard 
deviations away from the mean amplitude a positive density perturbation 
must lie to 
have collapsed and virialized by time $t$.  The total mass density in 
virialized perturbations of mass $M$ or greater is then proportional 
to the integral over the tail of this Gaussian distribution from 
$\nu_c$ to infinity, which yields the complementary error function in 
the above expression if $\sigma(M) \rightarrow 0$ as $M \rightarrow \infty$.
Because the typical amplitudes of perturbations decline with increasing
$M$, the number density of virialized objects exceeding mass $M$ is 
approximately $M^{-1} \rho(>M)$.  At several standard deviations away
from the mean, virialized objects with masses significantly exceeding $M$
are much rarer than objects of mass $M$, so this approximation for the
mean number density of virialized clusters is good to a few tens of 
percent.

We will now proceed to determine the expected number density of
clusters like MS1054-0321 at high redshift as follows:
\begin{enumerate}
\item We conservatively assume that MS1054-0321 is hotter than 10~keV.
\item In an $\Omega_0 = 1$ universe, $T_X \propto (1+z)$ for clusters
        of a given mass, so the expected temperature at $z=0$ of a 
        cluster as massive as MS1054-0321 is $> 5.5 \, {\rm keV}$.
\item Modifying the temperature-mass relation derived from simulations
        by Evrard \etal (1996) to reflect the mass within a region whose 
        density is 200 times the critical value, we find
\begin{equation}
  M_{\rm vir} \approx (1.8 \times 10^{15} \, h^{-1} \, M_\odot) \,
                  \left( \frac {kT_X} {10 \, {\rm keV}} \right)^{3/2} \; \; .
\end{equation}
        This relation yields a virial mass $> 7.4 \times 10^{14} \, 
        h^{-1} \, M_\odot$ for MS1054-0321.
\item According to Henry (1997) the current number density of clusters 
        hotter than $5.5 \, {\rm keV}$ is $< 3 \times 10^{-7} \, h^3 \,
        {\rm Mpc}^{-3}$.  This number density decreases rapidly to higher
        temperatures and masses, so the mean virialized mass density
        in clusters hotter than $5.5 \, {\rm keV}$ is $< 2 \times
        10^{8} \, h^2 \, M_\odot \, {\rm Mpc}^{-3}$.
\item Invoking the integral Press-Schechter formula, we find that
        $\nu_c(z=0) > 3.4$ for clusters with $T_X > 5.5$~keV.
\item Devolving the growth of perturbations back to $z=0.83$ implies
        that $\nu_c(z=0.83) > 6.2$ for clusters with $T_X > 10 \,
        {\rm keV}$.
\item At $z=0.83$, the mean virialized mass density of such clusters 
        in comoving coordinates is thus $< 3.1 \times 10^2 \, h^2 \, 
        M_\odot \, {\rm Mpc}^{-3}$, and the corresponding number 
        density is $< 4.4 \times 10^{-13} \, h^3 \, {\rm Mpc}^{-3}$.
\end{enumerate}
Next we will show that the detection of MS1054-0321 in the EMSS
implies a space density $\sim 10^{-8} \, h^3 \, {\rm Mpc}^{-3}$
for like clusters, over $10^4$ times what we predict from the
current cluster population and the assumption of Gaussian
perturbations in an $\Omega_0=1$ universe.  Before we go on, 
note that this test for $\Omega_0$ is independent of the mass
spectrum of perturbations because we are comparing clusters of
similar masses (e.g. Oukbir \& Blanchard 1997).

\subsection{Hot Clusters in the EMSS}

The EMSS is particularly useful for studying cluster evolution
because it has a quantifiable selection volume (see Henry \etal 
1992; H92 hereafter).  The usual procedure for determining the
number density of clusters detected in such a survey is to
compute $V_{\rm max}$, the maximum volume within which a cluster 
of a given flux could be detected, for each cluster. 
The number density of high-redshift clusters in the survey is then 
the sum of the $1/V_{max}$ for all clusters in the redshift shell of
interest.  Here we consider the $z=0.5-0.9$ shell, which
contains 4 EMSS clusters bright enough for ASCA to measure a 
temperature. 

Gioia \& Luppino (1994) give the cluster X-ray fluxes within the 
detection cell for the EMSS clusters at $z>0.5$. 
To determine $V_{\rm max}$ for each cluster, we calculate the 
maximum redshift $z_{\rm max}$ at which it could have been detected 
at each flux limit listed in H92, assuming that the cluster has a core 
radius of 0.25~Mpc and a $\beta-$model surface brightness 
distribution with $\beta = 2/3$, while correcting for the fraction 
of the cluster emission that falls into the EMSS detection cell 
(H92; Donahue, Stocke \& Gioia 1991).  This correction factor is 
slightly smaller than two for $z>0.5$ (H92).
We then integrate over 
$r^2 dr$ with $r = 2c H_0^{-1} [1 - (1+z)^{-1/2}]$ from 
the minimum redshift of the shell to either the maximum redshift 
of the shell or $z_{\rm max}$, whichever is smaller.  This computation 
gives the comoving volume per unit steradian associated with the cluster
within the specified redshift shell in a flat universe.  To determine
$V_{\rm max}$ we multiply this number by the solid angle corresponding to
the appropriate flux limit.  We compute $V_{\rm max}$ assuming both 
(a) no $K$-correction and (b) a $K$-correction with $f_\nu \propto 
\nu^{-0.5}$. Varying any of these assumptions within reasonable ranges 
changes the results by $\lesssim 10$\%.  

The $V_{\max}$ associated with MS1054-0321 in a flat universe is $6.9
\times 10^{7} \, h^{-3} \, {\rm Mpc}^3$ with no $K$-correction and $6.5
\times 10^{7} \, h^{-3} \, {\rm Mpc}^3$ with the $K$-correction.  These
volumes imply a comoving space density $\sim 1.5 \times 10^{-8} \, h^3
\, {\rm Mpc}^{-3}$ for clusters like MS1054-0321 in the redshift shell
$z = 0.5-0.9$.  Over four orders of magnitude separate this density
from the expected value $\sim 4 \times 10^{-13} \, h^3 \, {\rm Mpc}^{-3}$,
casting severe doubts on either the assumption of $\Omega_0 = 1$
or the assumption of Gaussian initial perturbations.

While this single hot high-$z$ cluster argues strongly against 
$\Omega_0 = 1$, it does not constitute an airtight case on its own.  
One could argue that MS1054-0321 is anomalous in some way, but there
are two other high-$z$ clusters in the EMSS that are nearly as hot:
MS0016+16 with $T_X = 8.4$~keV at $z=0.54$ (Yamashita 1994; Furuzawa 
\etal 1994) and MS0451-03 with $T_X = 10.4 \pm 1.2 \, {\rm keV}$ at
$z=0.54$ (Donahue 1996).  Figure~\ref{tx_sigma} demonstrates the consistency
between their X-ray temperatures and optical velocity dispersions. 
(MS1137+66, the fourth high redshift EMSS cluster, was observed by
ASCA in 1997, and will be discussed in a separate paper.)

For comparision, we can compute the expected number density 
for these clusters based on the
same assumptions as before.  Given $T_X > 8$~keV at $z = 0.5$, we find
$M_{\rm vir} > 6.9 \times 10^{14} \, h^{-1} \, M_\odot$, corresponding
to $T_X > 5.3$~keV at $z=0$.  The current mean virialized mass density 
in such clusters is similar to the case of MS1054-0321, implying
$\nu_c > 3.4$ at $z=0$ and $\nu_c > 5.2$ at $z=0.54$.  For $\Omega_0
=1$ and Gaussian perturbations, the expected number density of $> 8$~keV
clusters at $z=0.54$ is then $< 8.6 \times 10^{-11} \, h^3 \, {\rm Mpc}^{-3}$.
Each of the two hot EMSS clusters at $z=0.54$ would therefore be 
rarer than one in a hundred, with a joint improbability 
of order a few times $10^{-4}$, somewhat more probable than
the chance occurrence of MS1054-0321.

In summary, given the three hot high redshift clusters in the EMSS, the implied number density of clusters with $T_X > 8 \, {\rm keV}$ 
at $z =0.5-0.9$ is $\sim 1.6 \times 10^{-8} \, h^3 \, {\rm Mpc}^{-3}$. The
probability of all three clusters appearing in the volume sampled by
the EMSS in an $\Omega=1$ Universe is $\sim 10^{-6}$ if the initial
density perturbations have a Gaussian distribution.

Universes with $\Omega_0 \approx 0.3$ have no problem accomodating such
hot clusters at these redshifts.  Eke \etal 
(1996; see also Viana \& Liddle 1996) show that the
cluster temperature function evolves very little in an $\Omega_0 = 0.3$
universe with no cosmological constant and only modestly in a flat universe
with the same $\Omega_0$.  The number density of $> 8$~keV clusters
from Henry (1997) is $\sim 5 \times 10^{-8} \, h^3 \, {\rm Mpc}^{-3}$,
quite close to the value we find at $z \sim 0.5$.  The statistics
of hot clusters at $z > 0.5$ therefore add substantial weight
to the growing body of evidence emerging from cluster evolution
that $\Omega_0 \approx 0.3-0.5$ (e.g., Carlberg \etal 1997; Henry 1997;
Bahcall, Fan, \& Cen 1997).

\subsection{High-$z$ $L_X$-$T_X$ Relation}

Cluster evolution from $z \sim 0.5$ to the present should be
modest at best in a universe with low $\Omega_0$.  Here we show
that the relationship between the X-ray luminosity ($L_X$) and the
X-ray temperature of MS1054-0321 is similar to the relationship
derived from lower-redshift clusters.  Edge \& Stewart (1991b) 
found that $L_X \approx (8.8 \times 10^{41} \, h^{-2} \, {\rm erg 
\, s^{-1}})(T_X / 1 \, {\rm keV})^{2.79 \pm 0.05}$ in the 
$2-10$~keV band for low-redshift clusters, while Mushotzky \& Scharf
(1997) demonstrate that clusters observed at $z > 0.14$ by ASCA, 
including MS0451-03 (Donahue 1996) and MS0016+16 (Yamashita 1994; 
Furuzawa \etal 1994), do not deviate appreciably from this relation.
Our observations indicate that $L_X(2-10 \, {\rm keV}) \approx 7.7 \times
10^{44} \, h^{-2} \, {\rm erg \, s^{-1}}$ for MS1054-0321 in its own
rest frame ($q_0=0.1$), implying an expected
temperature of 11.3~keV, on the lower edge of our one-dimensional 
90\% confidence interval.

The lack of evidence for $L_X$-$T_X$ evolution in our observation
of MS1054-0321 is consistent with the other evidence for $\Omega_0
< 1$, but this test for $\Omega_0$ is not as powerful as the 
temperature function test, as discussed in Section 3.2.  
Some cluster evolution models predict
little evolution in the $L_X$-$T_X$ relationship in an $\Omega_0 = 1$
universe, as long as the minimum entropy of the X-ray gas remains 
the same for all clusters regardless of redshift (Kaiser 1991, 
Evrard \& Henry 1991, Bower 1997). The lack of evolution
in $L_X$-$T_X$ would then require some sort of non-gravitational
heating of the X-ray gas.

\section{Summary}

The ASCA observations we have presented here imply that the EMSS
cluster MS1054-0321 at $z=0.83$ has an X-ray temperature of 
$12.3^{+3.1}_{-2.1} \, {\rm keV}$.  
Supporting optical redshifts of 12 member galaxies (this paper)
and the weak lensing map (Luppino \& Kaiser 1997) 
provide a velocity dispersion and
a lensing mass that agree with the mass inferred from the 
high X-ray temperature.
Our ROSAT image exhibits significant substructure, but numerical
modelling of cluster evolution has shown that the presence of 
substructure does not preclude using a cluster's X-ray temperature 
to estimate its virial mass (e.g. Evrard \etal 1996).

This high a temperature in so distant a cluster is 
powerful evidence that $\Omega_0 < 1$.  Using the integral
Press-Schechter relation and the current temperature function
of clusters, we calculate the expected number density of
such hot clusters at $z = 0.83$, under the assumptions of
Gaussian perturbations and $\Omega_0 = 1$.  We find that
the product of the expected number density and the effective
volume of the EMSS in the $z = 0.5-0.9$ redshift bin is less 
than a few times $10^{-5}$.  Similar calculations for two
other hot EMSS clusters at $z \approx 0.54$ give probabilities
less than a few times $10^{-3}$ for each one.

For $\Omega_0 \approx 0.3$, the temperature function of hot 
clusters is expected to evolve very little to  
redshifts $>0.5$ (Eke \etal 1996), so that the hot EMSS clusters are
quite consistent with a 
low-$\Omega_0$ universe.  We also find that MS1054-0321 luminosity
and temperature are consistent 
with the low-$z$ $L_X$-$T_X$ relation, a result also consistent with 
$\Omega_0 < 1$.  These results strongly support the mounting
body of evidence (e.g. Carlberg \etal 1997) 
showing that the mean density of matter in the
universe is $\Omega_0 \approx 0.3$.

\acknowledgements MD acknowledges Timothy Beers for the use of his
ROSTAT program to calculate velocity dispersions, and helpful discussions
with Carlos Frenk and Bharat Ratra at the Aspen Center for Physics. This
work was supported by NASA ROSAT
grant NAG5-2021, NASA ASCA grant NAG5-2570. IMG acknowledges partial 
support from NASA STScI grant GO-542.01-93A and ASI94-RS-10. IMG acknowledges
support from NSF AST95-00515. 
JPH's research on clusters is supported by NASA Long Term Space
Astrophysics Grant NAG5-3432.  MD acknowledges Paul Lee for assistance 
with ASCA data extraction.

\newpage

\begin{table}[t]
\caption{ASCA Observing Information \label{asca} }
\begin{tabular}{lrrr}
Detector & Aperture Radius & Count Rates & Useful Exposure Time \\
    & Arcmin &  s$^{-1}$ & s \\ \tableline
SIS0  & 2.53  & 0.019 &  62,019 \\
SIS1 &  3.25  & 0.013 &  58,176 \\
GIS2 &  5.00 &  0.012 &  65,534 \\
GIS3 &  5.00 &  0.016 &  65,920 \\ \tableline
\end{tabular}
\end{table}

\begin{table}[h]
\caption[]{Member Galaxy Redshifts \label{Redshifts}}
\begin{tabular}{cccc}
ID &	Redshift  & Offset E/W (+/-") & Offset N/S (+/-") \\ \tableline
1 (cD) &    $0.8309 \pm 0.0008$ & 0.0 & 0.0 \\
2 &	$0.8342  \pm 0.003$     & +23 & -6 \\
3 &   $0.8127 \pm 0.0003$      & +56  & -5.5 \\
4 &   $0.8213 \pm 0.0007$      & +37  & +30 \\
5 &  $0.830: \pm 0.008$ 	& -14 & -20 \\
6 &  $0.8209 \pm 0.001$     	& -29 & -13 \\
7 &   $0.8286 \pm 0.001$ 	& -31.5 & -11.5 \\
8 &   $0.8353 \pm 0.0006$	& -38  & -8 \\
9 &   $0.8332 \pm 0.001$ 	& -43 & -6 \\
10 & 	$0.8143 \pm 0.002$ 	& -38 & +0.5 \\
11  &	$0.8378 \pm 0.003$ 	& -80 & -44 \\
12  &	$0.8319 \pm 0.002$ 	& -97.5 & -38.5 \\ \tableline
\end{tabular}
\end{table}

\newpage

\begin{figure}
\figcaption[plate1.ps]
{Plate 1. The central $3.75\arcmin \times 3.75\arcmin$ 
of the cluster MS1054-0321. The optical image is a 14,400 second I-band
image taken with the University of Hawaii 88-inch (Gioia \& Luppino 1994). The ROSAT
HRI image, a sum of all available HRI data for a total livetime of
121,590 seconds, 
was rebinned into $8\arcsec \times 8\arcsec$ 
pixels and adaptively smoothed with 
three Gaussians with sigmas of 16, 14.3, and 12.6 arcseconds corresponding
to counts in a pixel of $<10$, $10-13.5$ and $>13.5$. The sigmas
were chosen so that the products of the number of counts and $\sigma^2$ were
roughly constant. The background was 9.4164 counts per pixel.
The net contour levels are 1, 2.8, 4.6, 6.4, 8.2, and 10 counts per pixel.
These net count rates
correspond to X-ray surface brightnesses of  1.4, 3.9, 11.6, and
$14.0 \times 10^{-14}$ erg cm$^{-2}$ s$^{-1}$ arcmin$^{-2}$ assuming
that 1 HRI count $\sim 2.8 \times 10^{-11}$ erg cm$^{-2}$. The galaxies
are marked with identifications corresponding to Table~\ref{Redshifts}.
The central galaxy is at RA(2000)=10 56 59.9 and Dec(2000)=-03 37 37.3,
as measured from our HST image of this cluster (Donahue et al., in 
preparation). \label{plate}}
\end{figure}


\newpage
\begin{references}
\reference{a96} Arnaud, K. A. 1996, Astronomical Data Analysis Software
and Systems V, eds. G. Jacoby and J. Barnes, ASP Conf. Series vol 101.

\reference{bah97} Bahcall, N. A., Fan, X., \& Cen, R. 1997, \apj,
   490, L123.

\reference{beers90} Beers, T.C., Flynn, K., \& Gebhardt, K.
1990, \aj, 100, 32.

\reference{bower87} Bower, R. G. 1997, \mnras, 288, 355

\reference{car97} Carlberg, R. G., Morris, S. L., Yee, H. K. C., \&
   Ellingson, E. 1997, \apj, 479, L19.

\reference{car96} Carlberg, R. G., Yee, H. K. C., Ellingson, E.,
  Abraham, R., Gravel, P., Morris, S., \& Pritchet, C. J. 1996, \apj,
  462, 32




\reference{day} Day, C., Arnaud, K., Ebisawa, K., Gotthelf, E., Ingham, J.,
	Mukai, K., \& White, N. 1995, ``The ABC Guide to ASCA Data
	Reduction, Fourth Version'', available by request from the ASCA GOF.

\reference{d} Donahue, M. 1993 in Evolution of Galaxies and Their Environment,
(Kluwer: Dordrecht), eds. J. M. Shull \& H. A. Thronson, Jr., p. 409.

\reference{d96} Donahue, M. 1996, \apj, 468, 79.
\reference{dsg} Donahue, M., Stocke, J. T., \& Gioia, I. M. 1991, \apj, 385, 49.
 
\reference{es91a}  Edge, A. C. \& Stewart, G. C. 1991a, \mnras, 252, 428.

\reference{es91b} Edge, A.~C., \& Stewart, G.~C. 1991b, \mnras, 252, 414

\reference{eke} Eke, V. R., Cole, S., \& Frenk, C. S. 1996, MNRAS, 282, 263.

\reference{e89} Evrard, A. E. 1989, \apj, 341, L71

\reference{eh}  Evrard, A. E. \& Henry, J. P. 1991, \apj, 383, 95.

\reference{emn} Evrard, A. E., Metzler, C. A. \& Navarro, J. F. 
1996, \apj, 469, 494.


\reference{lee-stat} Fitchett, M. 1988, MNRAS, 230, 161.

\reference{furuzawa} Furuzawa, A., Yamashita, K., Tawara, Y., Tanaka, Y., 
        \& Sonobe, T. 1994, in New Horizon of X-ray Astronomy,
	eds. F. Makino and T. Ohashi, (Universal Academy Press: Tokyo), 
        p. 541.


\reference{gioia90b}  Gioia, I. M., Maccacaro, T., Schild, R. E., Wolter,
        A., Stocke, J. T. 1990, \apjs, 72, 567.

\reference{gl} Gioia, I. M. \& Luppino, G. A. 1994, \apjs, 94, 583.


\reference{henry}  Henry, J. P., Gioia, I. M., Maccacaro, T., Morris, S. L.,
        Stocke, J. T. 1992, \apj, 386, 408. (H92). 



\reference{h} Henry, J. P. 1997, \apj, 489, L1

\reference{hb} Hughes, J.~P., \& Birkinshaw, M.~1998, \apj, 
         in press (astro-ph/9711203) 

\reference{hbh} Hughes, J.~P., Birkinshaw, M., \& Huchra, J.~P.~1995,
         \apj, 448, L93

\reference{k91}  Kaiser, N. 1991, \apj, 383, 104.

\reference{lc94} Lacey, C., \& Cole, S. 1994, \mnras, 271, 676

\reference{lee} Lee, K. L. 1979, J. Am. Statistical Ass., 74, 708.


\reference{lk} Luppino, G. A. \& Kaiser, N. 1997, \apj, 475, 20.


\reference{ms} Mushotzky, R. \& Scharf, C. A. 1997, ApJ, 482, L13.



\reference{oke95} Oke, J.B., Cohen, J.G., Carr, M., Cromer, J., Dingizian, A., Harris, F.H., 
Labrecque, S., Luciano, R., Schaal, W., Epps, H., \& Miller, J., 1995,
PASP, 107, 375.

\reference{ob92} Oukbir, J. \& Blanchard, A. 1992, A\&A, 262, L21.
\reference{ob97} Oukbir, J. \& Blanchard, A. 1997, A\&A, 317, 1.


\reference{p} Peebles, P, J. E. 1993, Principles of Physical Cosmology, 
         Princeton: Princeton University Press, p. 98.

\reference{pdj} Peebles, P. J. E., Daly, R., \& Juszkeiewicz, R. 1989,
    \apj, 347, 563

\reference{ps} Press, W. \& Schechter, P. 1974, \apj, 187, 425.


\reference{roett} Roettiger, K., Burns, J. O., \& Loken, C. 1996, \apj, 
         473, 651.

\reference{schind} Schindler, S. 1996, \aap, 305, 756.


\reference{tih} Tanaka, Y., Inoue, H., \& Holt, S. S. 1994, \pasj, 46, L37.


\reference{vl} Viana, P. T. P. \& Liddle, A. R. 1996, MNRAS, 281, 323.


  
\reference{y} Yamashita, K. 1994,  in New Horizon of X-ray Astronomy,
eds. F. Makino and T. Ohashi, (Universal Academy Press: Tokyo), p. 279.

\end{references}
\end{document}